\documentclass{article}

\usepackage[cmfonts,ninepoint]{eusipco}  

\usepackage{psfrag}
\usepackage{amsmath}
\usepackage{amssymb}
\usepackage{amsfonts}
\usepackage{color}
\usepackage{epsfig}

\newcommand{\argmin}{\mathop{\mathrm{argmin}}}
\def\PSNR{\mathrm{ PSNR}}
\def\punit{\, \mathrm}

\title{Motion Compensated Frequency Selective Extrapolation for Error Concealment in Video Coding}
\name{J\"urgen~Seiler and Andr\'e~Kaup}
\address{Chair of Multimedia Communications and Signal Processing, \\University of Erlangen-Nuremberg, Cauerstr. 7, 91058 Erlangen, Germany\\
{\{seiler, kaup\}@LNT.de}}

\begin{document}

\maketitle

\begin{abstract}
Although wireless and IP-based access to video content gives a new degree of freedom to the viewers, the risk of severe block losses caused by transmission errors is always present. The purpose of this paper is to present a new method for concealing block losses in erroneously received video sequences. For this, a motion compensated data set is generated around the lost block. Based on this aligned data set, a model of the signal is created that continues the signal into the lost areas. Since spatial as well as temporal informations are used for the model generation, the proposed method is superior to methods that use either spatial or temporal information for concealment. Furthermore it outperforms current state of the art spatio-temporal concealment algorithms by up to 1.4 dB in PSNR.
\end{abstract}

\section{Introduction}

Due to modern video codecs and increased computational power, the transmission and processing of video signals in wireless environments or IP-based access to videos became more and more usual in the past years. Unfortunately, in these cases the risk of data lost in transmission or erroneously received is omnipresent. To cope with this risk, modern video codecs such as H.264/AVC use two strategies. According to \cite{Stockhammer2005}, the first one is error resilience by protecting the coded video against transmission errors and by minimizing the potential damage produced by incorrectly received bits. In the case that an error occurs, the second strategy has to take place, the concealment of block losses. Although the concealment is not part of the actual video standard, error concealment is widely used in many decoders in order to display a pleasant video and to reduce error propagation. A good overview over this area can be found in \cite{Wang1998}.

For concealing a lost block, most existent techniques use either spatial or temporal information for extrapolating the signal into the area of the lost block. The spatial methods only use information from the neighborhood of the lost block in the actual frame to extrapolate the signal into the area of the lost block. Two algorithms out of this group are e.\ g.\ the Projections onto Convex Sets from \cite{Sun1995} and the Sequential Error-Concealment from \cite{Li2002}. On the other hand, the temporal methods extrapolate the signal into the lost area only by means of information from previous or following already correctly received frames. In most cases, these methods try to estimate the motion in a sequence and replace the lost block with one from another frame, shifted according to the estimated motion. Two powerful temporal concealment algorithms are e.\ g.\ the Extended Boundary Matching Algorithm from \cite{Lam1993} or the Decoder Motion-Vector Estimation from \cite{Zhang2000a}. Unlike these two groups, spatio-temporal algorithms use spatial as well as temporal information for concealing the errors. One state of the art spatio-temporal concealment algorithm is the Three-Dimensional Frequency Selective Extrapolation (3D-FSE) introduced by \cite{Meisinger2007}. This algorithm aims at generating a model of the signal for a data volume centered by the erroneous block. The 3D-FSE is able to implicitly compensate small motion inside the volume, leading to very good objective and subjective extrapolation results. 

Now, we propose a new concealment technique, the Motion Compensated Frequency Selective Extrapolation (MC-FSE). It is based on the 3D-FSE but uses an explicit motion estimation and compensation prior to the model generation. By explicitly estimating the motion and aligning the extrapolation volume, significantly better extrapolation results can be obtained compared to 3D-FSE.

\section{Motion Compensated Frequency Selective Extrapolation}\label{sec:3d_fse}

\begin{figure*}
 	\centering
	\psfrag{x}[c][c][0.9]{$x$}
	\psfrag{y}[c][c][0.9]{$y$}
	\psfrag{t}[c][c][0.9]{$t$}
	\psfrag{m}[c][c][0.9]{$m$}
	\psfrag{n}[c][c][0.9]{$n$}
	\psfrag{p}[c][c][0.9]{$p$}
	\psfrag{A}[c][c][0.9]{$\mathcal{A}$}
	\psfrag{B}[c][c][0.9]{$\mathcal{B}$}
	\psfrag{P}[c][c][0.9]{$\mathcal{L}$}
	\psfrag{x0}[c][c][0.9]{$x_0$}
	\psfrag{y0}[c][c][0.9]{$y_0$}
	\psfrag{ttau}[c][c][0.9]{$t=\tau$}
	\psfrag{ttaum1}[c][c][0.9]{$t=\tau-1$}
	\psfrag{ttaup1}[c][c][0.9]{$t=\tau+1$}
	\psfrag{dtm1}[c][c][0.9]{\color{red} $\mathbf{d}^{\left(-1\right)}$}
	\psfrag{dtp1}[c][c][0.9]{\color{red} $\mathbf{d}^{\left(+1\right)}$}
	\psfrag{Video sequence}[l][l][1.0]{Video sequence}
	\psfrag{Motion compensated and}[l][l][1.0]{Motion compensated and}
	\psfrag{aligned extrapolation volume}[l][l][1.0]{aligned extrapolation volume}
 	\includegraphics[width=0.75\textwidth]{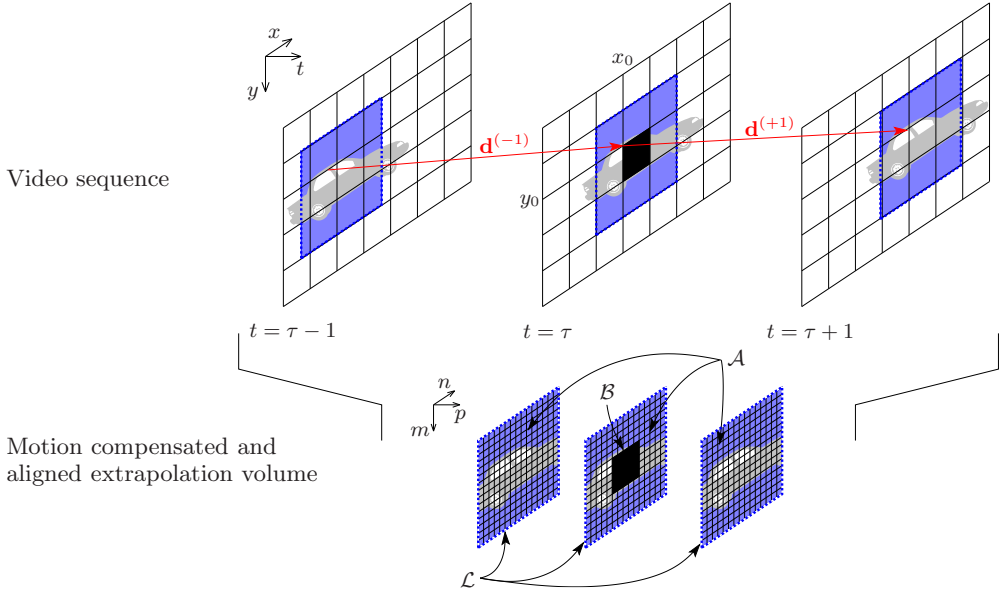}
 	\caption{Original video sequence $v\left[x,y,t\right]$ and aligned extrapolation volume $\mathcal{L}$. }
 	\vspace{-0.4cm}
 	\label{fig:extrapolation_area}
\end{figure*}

Fig.\ \ref{fig:extrapolation_area} shows three consecutive frames of a possible video sequence $v\left[x,y,t\right]$ depicted by the two spatial coordinates $x$ and $y$ and the temporal coordinate $t$. In the frame at time instance $t=\tau$ an isolated block loss occurs with the top left corner at $\left(x_0,y_0\right)$. Although isolated block losses are considered for the illustration of the algorithm, the algorithm easily can be adapted to other loss scenarios as well. In order to conceal the lost block, surrounding pixels in the actual frames and pixels in one or more previous and following frames are used. The number of used previous frames is indicated by $N_\mathrm{p}$, of used following frames by $N_\mathrm{f}$. In the case that the considered sequence is not a static one, the areas in the previous or following frame that correspond to the block loss may have been moved. Based on the frame at $t=\tau$, the movement relative to the frame at $t=\tau+\kappa$ is described by the motion vector 
\begin{equation}
\mathbf{d}^{\left(\kappa \right)}= \left(x^{\left(\kappa \right)}_\mathrm{d}; y^{\left(\kappa \right)}_\mathrm{d}\right)
 \end{equation}
whereas each motion vector includes the displacement in horizontal direction $x_\mathrm{d}^{\left(\kappa \right)}$ and in vertical direction $y_\mathrm{d}^{\left(\kappa \right)}$. For the actual concealment the motion has to be revoked in order to generate an aligned  three-dimensional data set, the so called extrapolation \mbox{volume $\mathcal{L}$}. The volume $\mathcal{L}$ is depicted by the spatial coordinates $m$ and $n$ and the temporal coordinate $p$. All in all, it is of size $M\times N\times P$. Further, it contains the lost block subsumed in area $\mathcal{B}$ and all the pixels used to extrapolate the signal into this area. These pixels are subsumed in volume $\mathcal{A}$, called support volume. 

\begin{figure}
 	\centering
 	\psfrag{Lost block}[c][c][0.9]{Lost block}
	\psfrag{Motion vector estimation}[c][c][0.9]{Motion vector estimation}
	\psfrag{Estimation reliability check}[c][c][0.9]{Estimation reliability check}
	\psfrag{Motion}[c][c][0.9]{Motion}
	\psfrag{vector}[c][c][0.9]{vectors}
	\psfrag{reliable?}[c][c][0.9]{reliable?}
	\psfrag{yes}[c][c][0.9]{yes}
	\psfrag{no}[c][c][0.9]{no}
	\psfrag{Volume alignment}[c][c][0.9]{Volume alignment}
	\psfrag{3D-FSE}[c][c][0.9]{3D-FSE}
	\psfrag{Cut out area B out of gmn}[c][c][0.9]{Cut out area $\mathcal{B}$ out of $g\left[m,n\right]$}
	\psfrag{Replace lost block}[c][c][0.9]{Replace lost block}
	\includegraphics[width=0.30\textwidth]{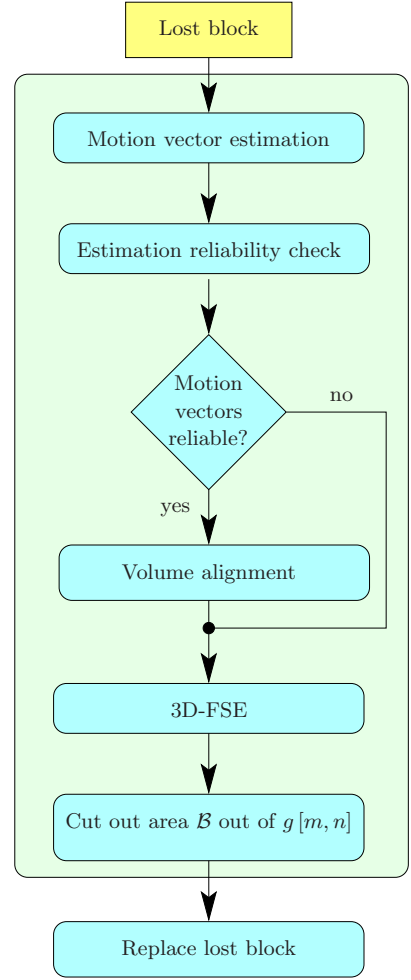}
	\caption{Block diagram for MC-FSE}
	\label{fig:block_diagramm}
\end{figure}

The block diagram in \mbox{Fig.\ \ref{fig:block_diagramm}} shows the different steps of the MC-FSE that will be carried out in detail in the subsequent subsections. Starting with the lost block, the motion of the sequence around this block is estimated. As this estimation may be inaccurate, the motion estimation's reliability is checked. If the motion estimation is reliable, the extrapolation volume is aligned according to the estimated motion. Afterwards, the Three-Dimensional Frequency Selective Extrapolation \cite{Meisinger2007} enhanced by the fast orthogonality deficiency compensation proposed in \cite{Seiler2008} is applied to the extrapolation volume. Finally, the area corresponding to the lost block is cut out of the generated model and is used for replacing the lost block.

\subsection{Motion estimation}\label{ssec:motion_estimation}

As mentioned above, the first step of the MC-FSE is the motion estimation for obtaining an aligned extrapolation volume. Although the 3D-FSE is able to inherently compensate minor motion, large motion cannot be compensated well as in this case the support volume covers inappropriate content of the previous and following frames. Thus, better extrapolation results are obtainable by aligning the extrapolation volume to revoke the motion.

\begin{figure}
 	\centering
	\psfrag{x0}[c][c][0.9]{$x_0$}
	\psfrag{y0}[c][c][0.9]{$y_0$}
	\psfrag{M}[c][c][0.9]{$\mathcal{M}$}
	\includegraphics[width=0.19\textwidth]{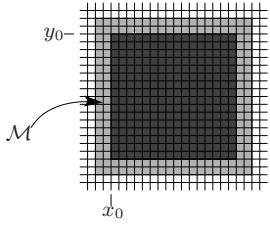}
	\caption{Decision area $\mathcal{M}$ for motion estimation of block at $\left(x_0,y_0\right)$}
	\vspace{-0.4cm}
	\label{fig:motion_estimation}
\end{figure}

Since the lost block cannot be used to determine the motion vector, the motion is estimated according to the Decoded Motion-Vector Estimation \cite{Zhang2000a}. Thereby, the estimation is performed by using an area around the lost block that here is called decision area $\mathcal{M}$ (see Fig.\ \ref{fig:motion_estimation}). For estimating the motion from the frame at $t=\tau$ to the frame at $t=\tau+\kappa$ a set of possible motion vectors is analyzed whereas for every possible motion vector $\tilde{\mathbf{d}}^{\left(\kappa \right)}= \left(\tilde{x}^{\left(\kappa \right)}_\mathrm{d}; \tilde{y}^{\left(\kappa \right)}_\mathrm{d}\right)$ the sum of squared errors 
\begin{equation}
E^{\left(\kappa \right)}\left(\tilde{\mathbf{d}}^{\left(\kappa \right)}\right) \hspace{-1mm} = \hspace{-4mm} \sum_{\forall \left(x,y\right) \in \mathcal{M}} \hspace{-4mm} \left(v\left[x,y,\tau\right] - v\left[x+\tilde{x}^{\left(\kappa \right)}_\mathrm{d}, y+\tilde{y}^{\left(\kappa \right)}_\mathrm{d}, \tau+\kappa\right] \right)^2
\end{equation}
is calculated for all pixels in area $\mathcal{M}$. After this, the motion vector is chosen that minimizes the sum of squared errors 
\begin{equation}
 \hat{\mathbf{d}}^{\left(\kappa \right)}  = \hspace{-5mm} \argmin_{\tilde{x}^{\left(\kappa \right)}_\mathrm{d}, \tilde{y}^{\left(\kappa \right)}_\mathrm{d} = -d_\mathrm{max}, \ldots, d_\mathrm{max}} \hspace{-5mm} E^{\left(\kappa \right)}\hspace{-1mm}\left(\tilde{x}^{\left(\kappa \right)}_\mathrm{d}, \tilde{y}^{\left(\kappa \right)}_\mathrm{d}\right).
\end{equation}
Here, the set of motion vectors to be tested covers all vectors with a maximum displacement of $d_\mathrm{max}$ in horizontal and vertical direction.

The estimation error $\check{E}^{\left(\kappa \right)}$ for the chosen motion vector is determined by
\begin{equation}
\check{E}^{\left(\kappa \right)} = E^{\left(\kappa \right)}\left( \hat{\mathbf{d}}^{\left(\kappa \right)}\right) .
\end{equation}

\subsection{Estimation reliability check}\label{ssec:estimation_check}

Unfortunately, in some cases the motion cannot be estimated well. This may happen e.\ g.\ in the event of non-translational movement or changes in the content. As no reliable motion vectors can be derived then, a compensated volume may be fitted less well for the extrapolation than an uncompensated one. To protect the algorithm from suffering from this circumstance, two thresholds are used to evaluate the motion vector quality. If at least one of these thresholds is exceeded, the motion vectors are discarded and no alignment is applied to the extrapolation volume.

The first threshold applies to the absolute motion vector quality. Therefore the maximum of the estimation error $\check{E}^{\left(\kappa \right)}$ for all preliminary calculated motion vectors is determined. This is normalized by the cardinality of $\mathcal{M}$ in order to get the mean error per pixel. The threshold for this criterion is denoted by $T_\mathrm{abs}$ and if
\begin{equation}
\frac{1}{\left|\mathcal{M}\right|} \max_{\kappa=-N_\mathrm{p},\ldots,N_\mathrm{f}\backslash 0} \hspace{-1mm} {\check{E}^{\left(\kappa \right)}}> T_\mathrm{abs}
\end{equation}
holds, all the estimated motion vectors are discarded.

The second criterion is the homogeneity of the motion vector quality. For this, the difference between the maximum estimation error and the minimum error is computed and normalized to the mean estimation error. The resulting quotient is compared to the threshold $T_\mathrm{rel}$. Thus, if
\begin{equation}
\frac{\displaystyle \max_{\kappa=-N_\mathrm{p},\ldots,N_\mathrm{f}\backslash 0} \hspace{-1mm} {\check{E}^{\left(\kappa \right)}} - \min_{\kappa=-N_\mathrm{p},\ldots,N_\mathrm{f}\backslash 0} \hspace{-1mm} {\check{E}^{\left(\kappa \right)}} }{ \displaystyle \frac{1}{N_\mathrm{p}+N_\mathrm{f}} \sum_{\kappa=-N_\mathrm{p},\ldots,N_\mathrm{f}\backslash 0} {\check{E}^{\left(\kappa \right)}} }> T_\mathrm{rel}
\
\end{equation}
is fulfilled, the motion vector quality is too distinct for getting a well aligned extrapolation volume. In this case, no alignment is applied and the volume is directly taken from the sequence without revoking any motion.

\subsection{Volume alignment}\label{ssec:volume_alignment}

After estimating the motion vectors for all frames used for the extrapolation, the projection volume $\mathcal{L}$ is set up. In the center, it consists of the lost block $\mathcal{B}$. This one is enframed by surrounding, correctly received pixels from the actual frame, for paying attention to the spatial neighborhood during the extrapolation process. From the used previous and following frames, the corresponding areas are shifted according to the appropriate motion vectors and are added to the extrapolation volume. The relation between the video sequence with the lost block and the aligned extrapolation volume is illustrated in Fig.\ \ref{fig:extrapolation_area} for the example of one previous and one following frame. So the motion of the sequence around the lost block is compensated and the aligned extrapolation is used for the subsequent Three-Dimensional Frequency Selective Extrapolation. 

\subsection{Frequency Selective Extrapolation}\label{ssec:fse}

After having performed the motion compensation for the extrapolation volume, the actual signal extrapolation can be carried out. It is based on the 3D-FSE \cite{Meisinger2007} combined with the orthogonality deficiency compensation proposed in \cite{Seiler2008}. In order to extrapolate the signal from the support volume $\mathcal{A}$ into the loss area $\mathcal{B}$ the algorithm aims at generating a parametric model $g\left[m,n,p\right]$ for the original signal. Therefore the model is generated by approximating the known signal in volume $\mathcal{A}$. As the model is defined over the complete volume $\mathcal{L}$, it continues the signal into area $\mathcal{B}$. The model
\begin{equation}
 g\left[m,n,p\right] = \sum_{\forall k\in\mathcal{K}} c_k \varphi_k \left[m,n,p\right]
\end{equation}
 itself is a weighted superposition of mutually orthogonal three-dimensional basis functions $\varphi_k \left[m,n,p\right]$. The expansion coefficients $c_k$ control the weight of each basis function. The set $\mathcal{K}$ covers the indices of all basis functions used for the extrapolation.

The model generation works iteratively whereas in every iteration step one basis function is chosen, denoted by index $u$. Then, this one is added to the parametric model from the previous iteration step together with an estimate $\hat{c}^{\left(\nu\right)}_u$ for the expansion coefficient. In the $\nu$-th iteration step, this leads to
\begin{equation}
 g^{\left(\nu\right)} \left[m,n,p\right] = g^{\left(\nu-1\right)} \left[m,n,p\right] + \hat{c}^{\left(\nu\right)}_u \cdot \varphi_u\left[m,n,p\right].
\end{equation}
 The initial model $g^{\left(0\right)}\left[m,n,p\right]$ is all zero. For all $\left(m,n,p\right) \in \mathcal{A}$ the residual approximation error can be computed according to
 \begin{equation}
  r^{\left(\nu\right)}\left[m,n,p\right] = r^{\left(\nu-1 \right)}\left[m,n,p\right] - \hat{c}^{\left(\nu\right)}_u \cdot \varphi_u\left[m,n,p\right].
 \end{equation}
As in area $\mathcal{B}$ no information about the original signal is existent, no approximation error can be computed there. The initial approximation error $r^{\left(0\right)}\left[m,n,p\right]$ is equal to the original signal in the volume $\mathcal{A}$.

In order to determine the basis function to use and to determine the estimate for the expansion coefficient, in every iteration step the projection coefficients $p_k^{\left(\nu\right)}$ are computed for all possible basis functions. $p_k^{\left(\nu\right)}$ results from the weighted projection of the approximation error onto the basis function $\varphi_k \left[m,n,p\right]$.
\begin{equation}
 p_k^{\left(\nu\right)} = \frac{\displaystyle \sum_{\left(m,n,p\right)\in \mathcal{L}} r^{\left(\nu-1\right)}\left[m,n,p\right] \cdot \varphi_k\left[m,n,p\right] \cdot w\left[m,n,p\right]}{\displaystyle \sum_{\left(m,n,p\right)\in \mathcal{L}} w\left[m,n,p\right] \cdot \varphi^2_k\left[m,n,p\right]}
\end{equation}
The weighting function 
\begin{equation}
 w\left[m,n,p\right] = \left\{ \begin{array}{lcl} \rho\left[m,n,p\right] &,& \forall \left(m,n,p\right) \in \mathcal{A} \\ 0 &,& \forall \left(m,n,p\right) \in \mathcal{B} \end{array} \right.
\end{equation}
is used to exclude area $\mathcal{B}$ from the projection process and to control the influence each pixel has on the extrapolation process depending on its position by means of $\rho\left[m,n,p\right]$.  According to \cite{Meisinger2007}, the weighting function used is generated by an isotropic model centered in the mid of the lost block. This leads to
\begin{equation}
 \label{eq:rho}
 \rho\left[m,n,p\right] = \hat{\rho}^{\sqrt{\left(m-\frac{M-1}{2}\right)^2+\left(n-\frac{N-1}{2}\right)^2+\left(p-\frac{P-1}{2}\right)^2}}
\end{equation}
with constant $\hat{\rho}$ between $0$ and $1$. So the pixels get less influence on the model generation with an increasing distance to the center of the lost block. After having computed all projection coefficients, the basis function to be added to the model is determined. The decision is made for the basis function that minimizes the distance between the residual error and the projection onto the basis function.

After having chosen one basis function the corresponding expansion coefficient has to be estimated. Although the basis functions are orthogonal with respect to the whole extrapolation volume $\mathcal{L}$, they are not orthogonal when evaluated with respect to the support volume $\mathcal{A}$. Due to this, the projection does not only lead to the real portion a basis function has on the approximation error but incorporates portions from other basis functions as well. In order to estimate the expansion coefficient $\hat{c}^{\left(\nu\right)}_u$ from the projection coefficient the fast orthogonality deficiency compensation proposed in \cite{Seiler2008} is used, resulting in
\begin{equation}
 \hat{c}^{\left(\nu\right)}_u = \gamma \cdot p_u^{\left(\nu\right)}.
\end{equation}
The orthogonality deficiency compensation factor $\gamma$ is constant and typically is in the range of $0$ to $1$.

The iteration steps are repeated until the maximum number of iterations is reached. Finally the pixels corresponding to the loss area $\mathcal{B}$ are cut out of $g\left[m,n,p\right]$ and are used to conceal the lost block.

\section{Simulation Setup and results}\label{sec:results}

\begin{figure*}
 	\centering 
 	\includegraphics[width=0.75\textwidth]{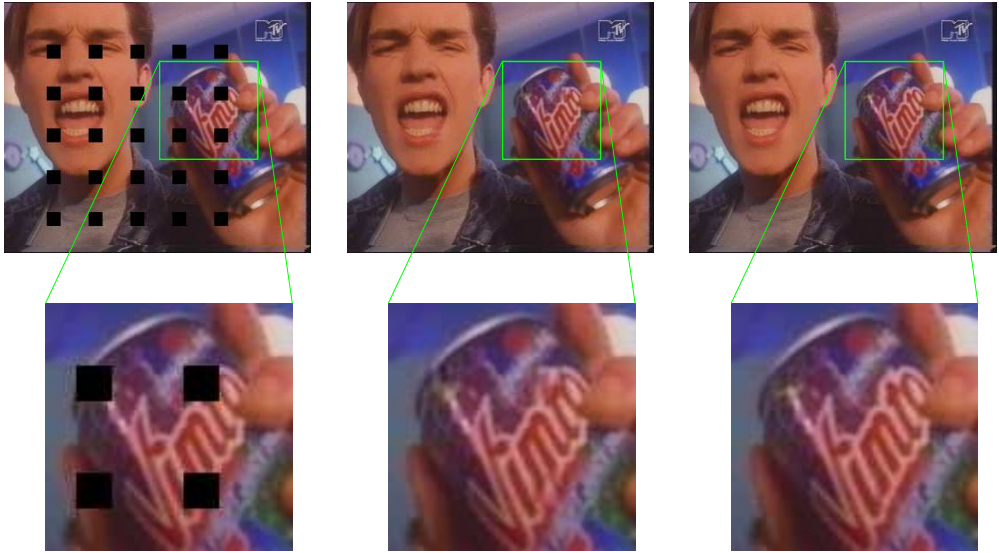}
	\caption{Visual extrapolation quality for frame $11$ of CIF sequence ``Vimto''. Left: Error pattern. Mid: 3D-FSE. Right: MC-FSE}
	\vspace{-0.4cm}
	\label{fig:vimto_concealed}
\end{figure*}

In order to demonstrate the extrapolation quality of the MC-FSE, the concealment of block losses in the CIF sequences ``City'', ``Foreman'', and ``Vimto'' is evaluated. Therefore, for all regarded sequences, in the frames number $17, 47, 77, 107,$ and $137$ blocks of size $16\times 16$ pixels are cut out according to the loss pattern showed on the left side of Fig.\ \ref{fig:vimto_concealed}. Then, for the luminance component these blocks are extrapolated and the extrapolation results are compared to the original blocks in terms of $\PSNR$.

The basis functions used for the extrapolation are the functions of the three-dimensional discrete Fourier transform. According to \cite{Meisinger2007}, this set of basis functions is especially suited for concealment of errors in natural images, since flat as well as noise like areas and edges can be reconstructed well. Additionally, an efficient implementation operating in the Fourier domain is possible \cite{Seiler2008}. For the actual extrapolation, $N_\mathrm{p}=2$ previous and $N_\mathrm{f}=2$ following frames are used. The support area in the actual frame is a band of $16$ pixels width around the lost block. This results in an extrapolation volume with an overall size of $48\times 48\times 5$ pixels. The decision area $\mathcal{M}$ for motion estimation is of $4$ pixels width and the search range is $d_\mathrm{max} = 16$ pixels in each direction at fullpel accuracy. As the efficient implementation operates in the Fourier domain, the extrapolation volume has to be transformed into this domain. Therefore, a FFT of size $64\times 64\times 16$ is used. As mentioned before, the weighting function is generated by an isotropic model. Regarding the variable $\hat{\rho}$ of the isotropic model, a value of $0.8$ has lead to good extrapolation results. For controlling the orthogonality deficiency compensation, the factor $\gamma$ is chosen to $0.6$ as this value is a good tradeoff between compensation quality and the number of iterations needed for achieving the maximum $\PSNR$. The two thresholds for evaluating the motion estimation quality are chosen to $T_\mathrm{rel}=3$ and $T_\mathrm{abs}=100$. Fortunately, these two thresholds as well as the values given before are not very critical and can be varied in a relatively wide range without affecting the extrapolation result very much. 

\begin{figure}
	\psfrag{s01}[t][t][0.9]{\color[rgb]{0,0,0}\setlength{\tabcolsep}{0pt}\begin{tabular}{c}Iterations\end{tabular}}%
	\psfrag{s02}[b][b][0.9]{\color[rgb]{0,0,0}\setlength{\tabcolsep}{0pt}\begin{tabular}{c}$\PSNR$ in $\punit{dB}$\end{tabular}}%
	\psfrag{s03}[b][b]{}%
	\psfrag{s06}[][]{\color[rgb]{0,0,0}\setlength{\tabcolsep}{0pt}\begin{tabular}{c} \end{tabular}}%
	\psfrag{s07}[][]{\color[rgb]{0,0,0}\setlength{\tabcolsep}{0pt}\begin{tabular}{c} \end{tabular}}%
	\psfrag{s13}[l][l][0.7]{\color[rgb]{0,0,0}``City'', 3D-FSE}%
	\psfrag{s14}[l][l][0.7]{\color[rgb]{0,0,0}``City'', MC-FSE}%
	\psfrag{s15}[l][l][0.7]{\color[rgb]{0,0,0}``Foreman'', 3D-FSE}%
	\psfrag{s16}[l][l][0.7]{\color[rgb]{0,0,0}``Foreman'', MC-FSE}%
	\psfrag{s17}[l][l][0.7]{\color[rgb]{0,0,0}``Vimto'', 3D-FSE}%
	\psfrag{s18}[l][l][0.7]{\color[rgb]{0,0,0}``Vimto'', MC-FSE}%
	\psfrag{x12}[t][t][0.9]{$0$}%
	\psfrag{x13}[t][t][0.9]{$100$}%
	\psfrag{x14}[t][t][0.9]{$200$}%
	\psfrag{x15}[t][t][0.9]{$300$}%
	\psfrag{x16}[t][t][0.9]{$400$}%
	\psfrag{x17}[t][t][0.9]{$500$}%
	\psfrag{x18}[t][t][0.9]{$600$}%
	\psfrag{x19}[t][t][0.9]{$700$}%
	\psfrag{x20}[t][t][0.9]{$800$}%
	\psfrag{x21}[t][t][0.9]{$900$}%
	\psfrag{x22}[t][t][0.9]{$1000$}%
	\psfrag{v12}[r][r][0.9]{$24$}%
	\psfrag{v13}[r][r][0.9]{$26$}%
	\psfrag{v14}[r][r][0.9]{$28$}%
	\psfrag{v15}[r][r][0.9]{$30$}%
	\psfrag{v16}[r][r][0.9]{$32$}%
	\psfrag{v17}[r][r][0.9]{$34$}%
	\psfrag{v18}[r][r][0.9]{$36$}%
 	\centering
 	\vspace{-0.4cm}
 	\includegraphics[width=0.48\textwidth]{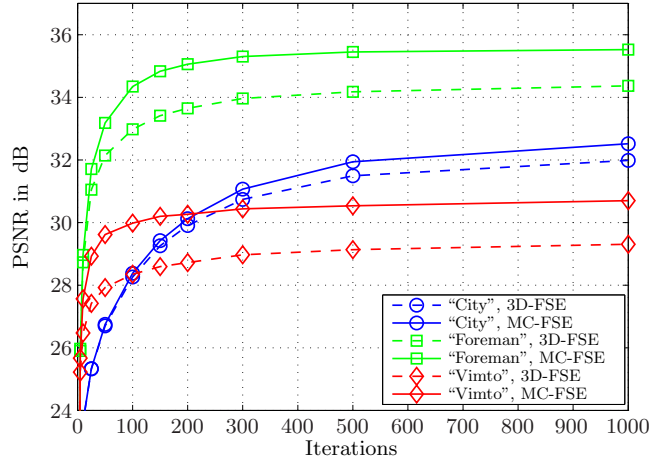}
	\caption{$\PSNR$ over iterations for 3D-FSE and MC-FSE. Block of size $16\times16$ pixels, $N_\mathrm{p}=2$, $N_\mathrm{f}=2$, $16$ pixels support area width, $4$ pixels wide $\mathcal{M}$, FFT $64\times64\times16$, $\hat{\rho}=0.8$, $\gamma=0.6$, $d_\mathrm{max}=16$, $T_\mathrm{rel}=3$, $T_\mathrm{abs}=100$}
	\vspace{-0.1cm}
	\label{fig:psnr_over_iter}
\end{figure}

In Fig.\ \ref{fig:psnr_over_iter}, for the mentioned sequences the obtainable $\PSNR$ is shown with respect to the number of iterations used for the model generation. For illustrating the gain of MC-FSE over 3D-FSE the graph also shows the $\PSNR$ for the 3D-FSE that performs the extrapolation on the non-aligned extrapolation volume. Except for the motion compensation, all other parameters are chosen in the same way for 3D-FSE as for MC-FSE. Thus, the orthogonality deficiency compensation from \cite{Seiler2008} is also applied to the 3D-FSE, although originally not used in \cite{Meisinger2007}. Obviously, by applying an explicit motion compensation and alignment of the extrapolation volume prior to the Frequency Selective Extrapolation the extrapolation quality can be increased significantly. For the considered sequences a maximum increment in $\PSNR$ of up to $1.4\punit{dB}$ is possible. Furthermore, most of the maximum gain already is existent at low numbers of iterations. Although the difference in $\PSNR$ depends on the sequence, for all tested sequences the explicit motion compensation leads to better extrapolation results.

\begin{table}
	\centering
		\begin{tabular}{|l|c|c|c|}
		\hline & ``City'' & ``Foreman'' &``Vimto'' \\
		\hline \hline TR & $25.63 \punit{dB}$ & $27.60 \punit{dB}$ & $22.79 \punit{dB}$ \\
		\hline EBMA \cite{Lam1993} & $22.53 \punit{dB}$ & $30.10 \punit{dB}$ & $28.14 \punit{dB}$ \\
		\hline DMVE \cite{Zhang2000a} & $27.94 \punit{dB}$ & $33.04 \punit{dB}$ & $28.21 \punit{dB}$ \\ 
		\hline 3D-FSE \cite{Meisinger2007} & $31.99 \punit{dB}$ & $34.37 \punit{dB}$ & $29.31 \punit{dB}$ \\
		\hline MC-FSE & $32.52 \punit{dB}$ & $35.53 \punit{dB}$ & $30.70 \punit{dB}$ \\ \hline
		\end{tabular}
	\caption{Comparison of different concealment algorithms}
	\vspace{-0.4cm}
	\label{tab:comparison}
\end{table}

Additionally, the MC-FSE is compared to several existent concealment techniques. As, according to \cite{Bopardikar2005}, in general temporal concealment algorithms are superior to spatial ones, the comparison is carried out to temporal methods only. The simplest temporal method for concealing a block loss is the Temporal Replacement (TR). Thereby, the lost block is replaced by the block from the previous frame that has the same position. A more sophisticated approach is the Extended Boundary Matching Algorithm (EBMA) from \cite{Lam1993} that replaces the lost block with a block from the previous frame that minimizes the boundary error between the correctly received neighboring blocks and the candidate block. Another method for concealing lost blocks is the Decoder Motion-Vector Estimation (DMVE) from \cite{Zhang2000a} that already served as basis for the motion estimation in Section \ref{ssec:motion_estimation}. The search range for EBMA and DMVE is set to $d_\mathrm{max}=16$ pixels as well. In Table \ref{tab:comparison} the obtainable $\PSNR$ is listed for the mentioned reference algorithms, the 3D-FSE \cite{Meisinger2007}, and the proposed Motion Compensated Frequency Selective Extrapolation (MC-FSE). For 3D-FSE and MC-FSE the number of previous and following frames is chosen to $N_\mathrm{p} = 2, N_\mathrm{f}=2$. In addition, for 3D-FSE and MC-FSE Table \ref{tab:frame_comparison} shows the concealment quality when different numbers of previous and following frames are used for the model generation. Apparently, due to making use of the spatial information as well, the 3D-FSE and the MC-FSE outperform the pure temporal concealment techniques. But, as already shown in Fig.\ \ref{fig:psnr_over_iter}, the explicit motion compensation of the extrapolation volume leads to better extrapolation results for the proposed algorithm.

Besides the objective evaluation of the extrapolation quality in terms of $\PSNR$, the subjective visual extrapolation quality is important. Although the 3D-FSE is able to conceal the losses without almost any visible artifacts, in some rare cases, the original signal cannot be reconstructed sufficiently. As already indicated by the $\PSNR$, the MC-FSE is superior to the 3D-FSE and additionally is able to conceal these critic losses better. Fig.\ \ref{fig:vimto_concealed} is used to illustrate such a case. The middle image is concealed with 3D-FSE and shows some minor artifacts, e.\ g.\ at the top of the ``t'' or the bottom of the ``m''. On the right side the same image is concealed with MC-FSE. With this new technique, the block losses can be concealed very good, and no artifacts are visible any more.

\begin{table}
\vspace{0.25cm}
	\centering
		\begin{tabular}{|l|c|c|}
		\hline & 3D-FSE \cite{Meisinger2007} &  MC-FSE \\ \hline
		$N_\mathrm{p} = 1, N_\mathrm{f}=0$ & $29.21 \punit{dB}$ & $29.35 \punit{dB}$\\ \hline
		$N_\mathrm{p} = 2, N_\mathrm{f}=0$ & $28.80 \punit{dB}$ & $30.01 \punit{dB}$\\ \hline
		$N_\mathrm{p} = 1, N_\mathrm{f}=1$ & $30.63 \punit{dB}$ & $31.08 \punit{dB}$ \\ \hline
		$N_\mathrm{p} = 2, N_\mathrm{f}=2$ & $29.31 \punit{dB}$  & $30.70 \punit{dB}$\\ \hline
		\end{tabular}
	\caption{Comparison of different numbers of frames used for concealment for sequence ``Vimto''}
	\vspace{-0.4cm}
	\label{tab:frame_comparison}
\end{table}

\section{Conclusion}\label{sec:conclusion}

The proposed algorithm is an enhancement to the already existent Three-Dimensional Frequency Selective Extrapolation. By explicitly compensating the motion of a sequence prior to the actual extrapolation, the potential of this powerful algorithm can better be tapped. With that, a very high subjective as well as objective extrapolation quality can be achieved. Demonstrated for concealment of isolated block losses, the proposed Motion Compensated Frequency Selective Extrapolation leads to an almost perfect reconstruction of the signal in the lost areas. Nevertheless, further work will focus on performing the motion estimation with fractional-pel accuracy and on reducing the complexity of the algorithm by selecting several basis functions in an iteration step. In addition to that, an implementation into a state-of-the-art video decoder is planned in order to evaluate the performance of the proposed algorithm in combination with more complex loss patterns in a realistic scenario. Although the MC-FSE is introduced for concealment of block losses it can as well be used for other signal extrapolation tasks as \mbox{e.\ g.\ } the prediction in video coding.




\end{document}